\documentclass[twocolumn]{aastex701} 
\usepackage{booktabs}
\usepackage{hyperref}
\usepackage{graphicx}
\usepackage{amsmath}
\usepackage{multirow}

\begin{document}

\title{Pre-Virialized Assembly at Cosmic Dawn: The Dynamics and Extreme Ionization of Compact Group CGG-z7 at $z\sim7.04$}
\author[0009-0007-9647-9907]{Xiaoyang Wei}
\affiliation{Department of Astronomy, Tsinghua University, Beijing 100084, China}
\email{xy-wei25@mail.tsinghua.edu.cn}

\author[0000-0001-8467-6478]{Zheng Cai}
\affiliation{Department of Astronomy, Tsinghua University, Beijing 100084, China}
\email[show]{zcai@mail.tsinghua.edu.cn}

\author[0000-0002-3489-6381]{Fujiang Yu}
\affiliation{Department of Astronomy, Tsinghua University, Beijing 100084, China}
\email{yufj@mail.tsinghua.edu.cn}

\author[0000-0001-6251-649X]{Mingyu Li}
\affiliation{Department of Astronomy, Tsinghua University, Beijing 100084, China}
\email{lmy22@mails.tsinghua.edu.cn}

\author[0000-0003-0111-8249]{Yunjing Wu}
\affiliation{Kavli Institute for the Physics and Mathematics of the Universe (WPI), The University of Tokyo Institutes for Advanced Study, The University of Tokyo, Kashiwa, Chiba 277-8583, Japan}
\email{yunjing.wu@ipmu.jp}

\correspondingauthor{\href{mailto:zcai@tsinghua.edu.cn}{Zheng Cai}}

\begin{abstract}

We report the discovery of CGG-z7, the most compact galaxy group at $z\gtrsim7$ identified to 
the north of the GOODS-North field, observed by the JWST POPPIES program. 
The system consists of at least six members within a projected size of $7.8\times5.7$ kpc$^2$, four of which are spectroscopically confirmed via [O {\sc iii}] and H$\beta$ emission. The group exhibits a low line-of-sight velocity dispersion ($\approx93.7$ km s$^{-1}$) 
relative to its substantial stellar mass ($M_* \approx 10^{9.8} M_{\odot}$), yielding a stellar-to-dynamical mass ratio of $M_*/M_{\rm{vir}} \approx 0.15$. 
This ratio, exceeding typical values for virialized halos by a factor of $3$, 
indicates that the system is highly likely 
not in dynamical equilibrium. Instead, we interpret 
CGG-z7 as a pre-virialized structure, likely 
a major merger caught near apocenter—capturing the rapid, chaotic formation 
of a massive ``Red Nugget". Spectroscopic analysis reveals extreme 
ionization conditions and low metallicity across the group. In particular, the central galaxy reaches an extraordinarily high [O {\sc iii}]/$H\beta$ ratio of $\sim18$, which is likely indicative of an obscured AGN.
CGG-z7 thus serves as a unique laboratory 
for the physics of 
pre‑virialized galaxy assembly, bridging the gap between 
turbulent high-$z$ assembly and the quiescent galaxies seen at cosmic noon.

\end{abstract}



\section{Introduction}
Compact galaxy groups are predicted to be a crucial transition stage—the immediate precursors to massive, coalesced systems \citep{2015ARA&A..53...51S,2023A&A...670L..11J,2025A&A...694A.218B,2025arXiv251113650S}. Critically, such dense structures within larger overdense regions (protoclusters) are predicted to account for a dominant fraction of the cosmic star formation activity at high redshift \citep[e.g.,][]{2013ApJ...779..127C,2017ApJ...844L..23C,2024MNRAS.528.6934S}, making them key laboratories for early galaxy assembly. Within such environments, galaxies experience frequent gravitational interactions and mergers, driving their rapid dynamical evolution \citep[e.g.,][]{1997ARA&A..35..357H,2017ApJ...844L..23C,2025MNRAS.536.2000L,2025arXiv250706284W,2025arXiv251011770F,2026ApJ...999...40L}. 

Dense, pre-virialized structures represent a highly accelerated ``fast-track" evolutionary pathway. 
While not all high-redshift overdensities are deterministically bound for immediate quiescence, the extreme kinematics within compact groups frequently trigger merger-driven compaction \citep{2014MNRAS.438.1870D,2016MNRAS.457.2790T}. During this process, tidal torques efficiently strip angular momentum, driving massive inflows of cold gas into galactic centers. This dissipative contraction forms ultra-dense, star-forming cores \citep{2014MNRAS.438.1870D,2023MNRAS.522.4515L}.

Building on this compaction phase, frequent mergers and tidal forces continue to shape the gas distribution: while inflows funnel gas toward the center, outflows and stripping simultaneously remove gas from the galaxy outskirts,
potentially triggering and fueling active galactic nuclei (AGN), while simultaneously accelerating gas consumption or ejection \citep[e.g.,][]{2011MNRAS.415.1027H,2014ApJ...789..112C,2025arXiv251014743P,2025A&A...704A.101S,2026ApJ...997..208S}. These processes make compact groups plausible progenitors of the dense, quiescent galaxies observed at later cosmic epochs \citep[e.g.,][]{2022MNRAS.513.3252A,2024A&A...683L...4J,2024ApJ...963...49K}. In this picture, the fast-track compaction followed by AGN-driven gas depletion naturally bridges rapid assembly with subsequent quenching.

Despite the theoretical necessity of this fast-track pathway, state-of-the-art cosmological simulations continue to underpredict the abundance of massive quiescent galaxies at $z>4$ by an order of magnitude \citep[e.g.,][]{2019A&A...632A..80G,2023Natur.619..716C,2025NatAs...9..280D}.
This tension persists largely because direct observational constraints on the pre-virialized merging and saturation phases during the epoch of reionization remain elusive. While several high-redshift compact groups have been identified \citep[e.g.,][]{2018Sci...362.1034D,2022A&A...665L...7S,2023A&A...670L..11J,2024A&A...683L...4J,2024A&A...684A.196Z,2024A&A...690A..55S,2025arXiv251113650S,2025A&A...694A.218B,2025arXiv251011770F}, capturing a system precisely at the tipping point of dynamic instability and extreme ionization at $z>7$ has remained out of reach.
With its unique combination of sensitivity and spatial resolution, JWST finally enables us to identify such a transforming progenitor.

In this paper, we report the discovery of CGG-z7, a profoundly compact, dynamically young galaxy group at $z\sim7.04$ revealed by JWST/NIRCam. By presenting extreme ionization conditions and non-virialized kinematics, CGG-z7 offers unprecedented direct evidence of saturation-phase assembly prior to coalescence and quenching. The structure of the paper is as follows. In Section~\ref{sec:data}, we describe the observations and data reduction. Section~\ref{sec:results} details our analysis of the group's morphology, kinematics, stellar populations, and emission-line properties. Finally, in Section~\ref{sec:summary}, we provide a summary of our findings and discuss their implications. We adopt $H_0=73~\mathrm{km~s^{-1}~Mpc^{-1}}$, $\Omega_m=0.27$ and $\Omega_\Lambda=0.73$ in our analysis. We use line wavelengths 4960.293\AA~ for [O {\sc iii}]$\lambda4959$, 5008.239\AA~ for [O {\sc iii}]$\lambda5007$ and 4862.670\AA~ for H$\beta$ based on the CHIANTI Database \citep{1997A&AS..125..149D,2021ApJ...909...38D}\footnote{https://db.chiantidatabase.org/}.

\section{Data Reduction}\label{sec:data}
The data in this paper come from the POPPIES program \citep{2025AAS...24620909K} in Cycle 3 of JWST (PID 5398). Our target galaxy CGG-z7 is a compact galaxy group centered at R.A. 189.17582, Dec. 62.34485, located to the north of the GOODS-North field \citep{2004ApJ...600L..93G}. POPPIES (The Public Observation Pure Parallel Infrared Emission-Line Survey) is a 400-hour purely parallel survey covering $\sim1455\ \text{arcmin}^2$. It takes grism spectra with $1-3$ filters (F444W, then F322W2/F277W+F356W), as well as images in 3-8 filters over 150 fields, providing a blind emission line survey. F444W, F200W, F115W images and F444W grism spectra are available for CGG-z7. We note that CGG-z7 lies more than 0.8 arcmin away from the boundary of the conventional GOODS-North footprint, placing it outside the deep-field coverage where additional multi-band data are available. Although our target is also covered by HST/ACS and Spitzer/IRAC, these data were not included in the analysis owing to their low signal-to-noise ratios and/or poor angular resolution.

The image reduction is based on the standard \texttt{jwst} pipeline \citep{Bushouse2025} version $1.18.1$ with reference file \texttt{jwst\_1364.pmap}. We begin the reduction on the public uncal files from MAST\footnote{\url{https://mast.stsci.edu/portal/Mashup/Clients/Mast/Portal.html}}. 
We performed astrometric calibration by registering the images to \textit{Gaia} \citep{2023A&A...674A..22G}. The photometry catalog was then generated from the calibrated images using \texttt{photutils} \citep{larry_bradley_2025_14889440}.
The final mosaic images are drizzled with \texttt{pix\_frac}=1.0 and a pixel size of $0.03''$.
The total exposure times of F444W, F200W, F115W are 1804, 8761, 1804 seconds, respectively.

The F444W grism spectra of CGG-z7 are reduced following the routine outlined by \citet{2023ApJ...953...53S,sun2024nircam_grism}\footnote{The codes and calibration data are available at \url{https://github.com/fengwusun/nircam_grism/}}. We refer interested readers to the SAPPHIRES early data release paper \citet{2025arXiv250315587S} for further details.

\begin{figure*}[htbp]
    \centering
    \includegraphics[width=0.96\linewidth]{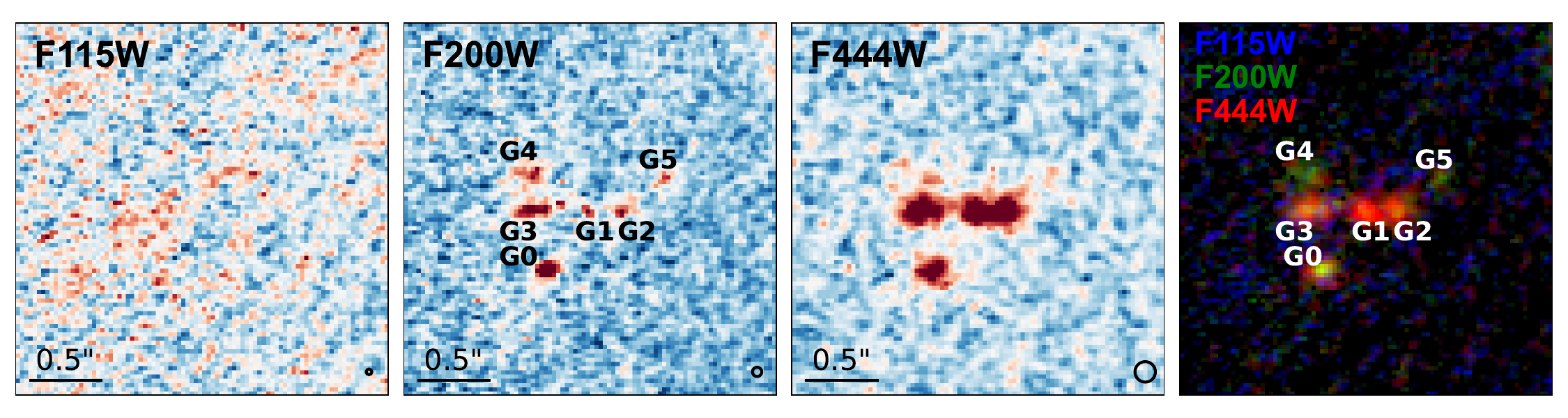}
    \caption{Multiband image of CGG-z7 (zscale). The black bar in the bottom left marks a $0.5''$ scale (2.59 kpc). The black circle in the bottom right denotes the FWHM of PSFs.}
    \label{fig:multiband}
\end{figure*}

\begin{figure*}[htbp]
    \centering
    \includegraphics[width=0.6\linewidth]{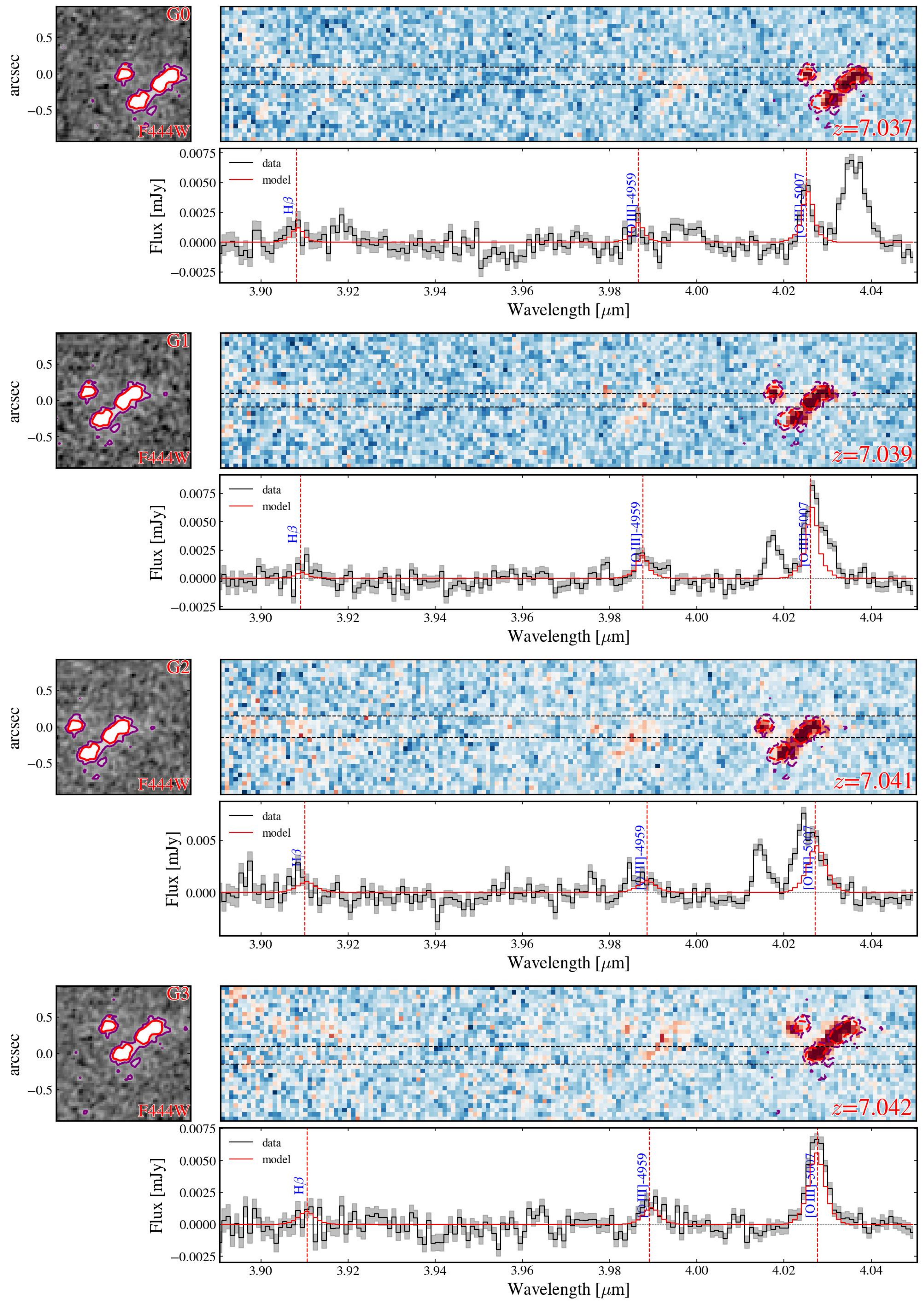}
    \caption{NIRCam F444W grism spectra of G0--G3 obtained through Cycle 3 PID 5398 (POPPIES). The F444W images here are rotated the observed PA of $218^\circ$ and flipped to match the dispersion direction of Module B of the JWST/NIRCam. For each source, clear detections of the H$\beta$ and [O {\sc iii}] doublet emission lines are visible. The 3$\sigma$ and 5$\sigma$ source contours extracted from the direct image are overlaid onto the 2D spectra. These contours have been translated to the spectral plane using the measured spectroscopic redshift and are centered at the expected location of the [O {\sc iii}]$\lambda$5007 line.}
    \label{fig:spec}
\end{figure*}

\section{Results}\label{sec:results}

\subsection{Morphology}\label{sec:morphology}
\begin{figure*}[htbp]
    \centering
    \includegraphics[width=0.96\linewidth]{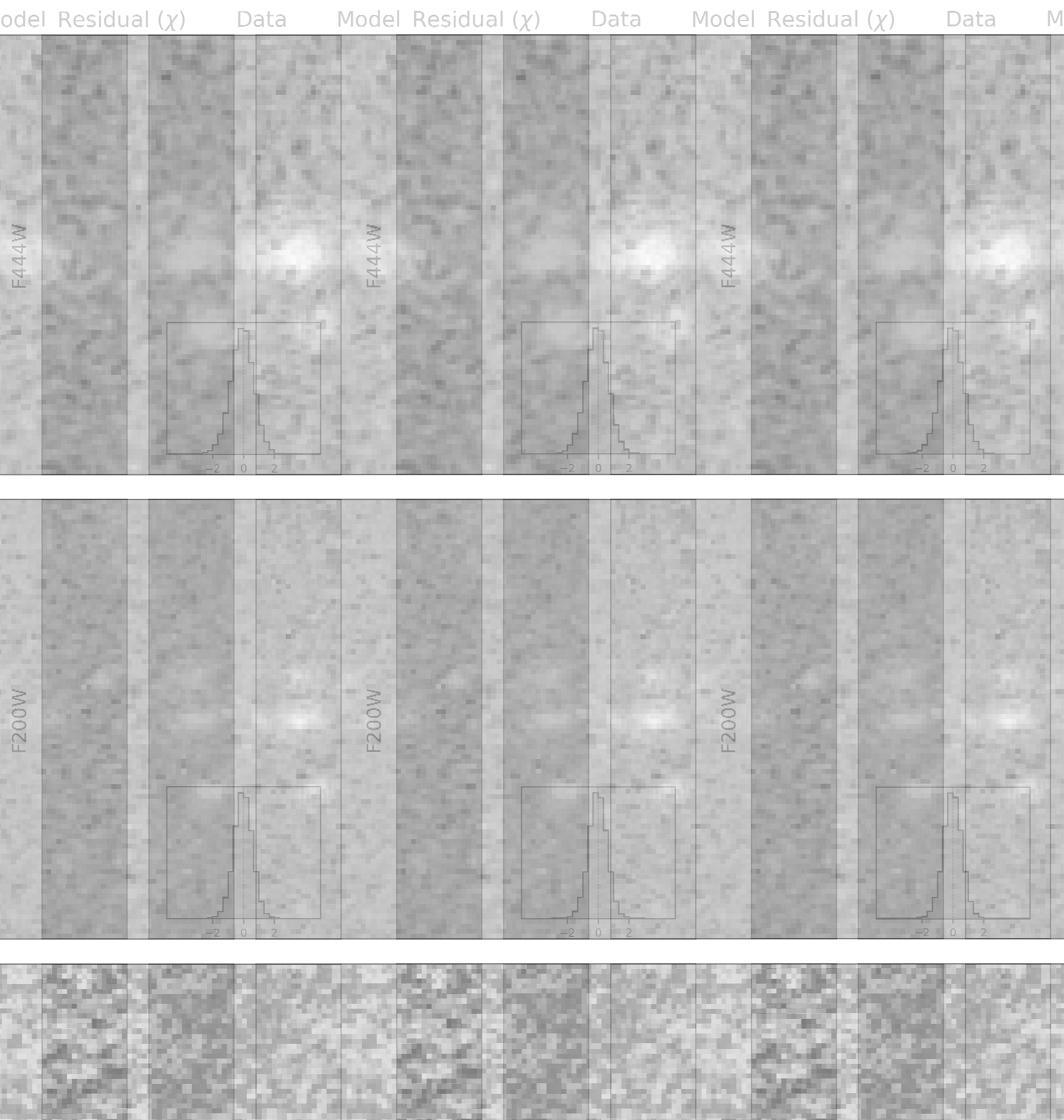}
    \caption{Multi-band morphological analysis of CGG-z7. The structure of the compact galaxy group CGG-z7 is modeled using \texttt{pysersic} with a multi-band joint fitting technique. Each row displays the results for one bandpass (F444W, F200W, and F115W, from top to bottom). Columns show, from left to right: the science image, the best-fit model, and the residual ($\chi=(\mathrm{data}-\mathrm{model})/\sigma_{\mathrm{data}}$) map. The inset in each residual panel shows the normalized histogram of $\chi$.}
    \label{fig:morphology}
\end{figure*}

This section presents the morphology analysis of CGG-z7 based on JWST/NIRCam imaging. As shown in Fig.~\ref{fig:multiband}, six candidate member galaxies are clearly detected in the deep F200W exposure, four of which are spectroscopically confirmed (see Sec.~\ref{sec:specz}). The group spans a projected area of approximately $1.5 \times 1.1$ arcsec$^2$, corresponding to $\sim 7.8 \times 5.7$ kpc$^2$ at its redshift.
We characterize the structural properties of the member galaxies using \texttt{pysersic} \citep{2023JOSS....8.5703P}. The analysis proceeds as follows. First, a detection map is created from the F200W image using \texttt{photutils} to identify the centroids of all group members. This map is used to generate a multi-band photometric catalog, which subsequently informs the fitting priors for \texttt{pysersic}. We then perform multi-band joint modeling, adopting a single Sérsic profile for each source in each band. During the fitting, we fix \texttt{xc,yc,theta} across the bands, and set \texttt{n,ellip,r\_eff} as the linked parameters\footnote{See \url{https://pysersic.readthedocs.io/en/latest/multiband-example.html}}.  A summary of the analysis is shown in Fig.~\ref{fig:morphology}. The marginalized posterior of the model parameters is shown in Tab.~\ref{tab:morphology_fit}. All of the members can be well fitted with a single Sérsic model. During the fitting, we use the empirical PSFs from \texttt{stpsf} \citep{Perrin2025}.

The morphological analysis yields Sérsic indices ranging from $n\sim$ 0.8 to 3.4, with galaxies G0, G2, and G3 exhibiting disky or irregular profiles ($n<2$). In the dense environment, such low-$n$ structures are unlikely to survive multiple strong gravitational encounters if the system were old and dynamically relaxed. Thus, the $n$'s provide indirect evidence that CGG-z7 is a dynamically young system, consistent with the kinematic picture derived from the stellar-to-dynamical mass ratio discussed in Sec.~\ref{sec:stellar}. The diversity in Sérsic indices among the group members suggests that these galaxies may have followed different evolutionary paths prior to their assembly into the current compact configuration. Nevertheless, all members share a common characteristic of being compact ($r_\mathrm{eff}\lesssim 0.8$ kpc). This combination of morphological diversity within a spatially concentrated, kinematically young system provides an interesting snapshot of early galaxy assembly, where galaxies with varied formation histories are drawn together into a single, evolving structure.

The morphology of G1 is particularly notable, as it is the most compact source in the group. Its best-fit \texttt{r\_eff}s of Sérsic profile are $\sim0.056$ arcsec in F444W and $\sim0.046$ arcsec in F200W, while its $3\sigma$ isophotal radii (defined as $\sqrt{\mathrm{area}/\pi}$ from the residual image after subtracting models of other members) are $\sim0.106$ arcsec in F444W and $\sim0.036$ arcsec in F200W. These sizes are comparable to or only slightly larger than half of the empirical PSF FWHM values (0.073 arcsec for F444W, 0.033 arcsec for F200W\footnote{\url{https://jwst-docs.stsci.edu/jwst-near-infrared-camera/nircam-performance/nircam-point-spread-functions}}), indicating that G1 remains largely unresolved and is likely a compact object. Its nature and implications will be further discussed in the following sections.

\subsection{Redshift}\label{sec:specz}
 As shown in Figure~\ref{fig:spec}, galaxies G0--G3 are spectroscopically confirmed using the combination of the [O~\textsc{iii}] doublet and H$\beta$ lines, whereas G4 and G5 are too faint to be detected in the grism spectra. The spectroscopic redshift $z_{\mathrm{spec}}$ for each confirmed galaxy is derived by cross-correlating a line-emission template with the extracted 1D spectrum. We note that the measured redshift is a combination of the cosmological redshift $z_{\mathrm{cos}}$ and the Doppler redshift $z_{\mathrm{pec}}$ due to peculiar velocity along line-of-sight (LoS), i.e.
 
 \begin{align}
     1+z_{\mathrm{spec},i}&=(1+z_{\mathrm{cos},i})(1+z_{\mathrm{pec},i})\\
     &\approx 1+z_{\mathrm{cos},i}+(1+z_{\mathrm{cos},i})v_i/c
 \end{align}
 where $v_i$ is the peculiar velocity along the LoS.    
The approximation holds for $v_i \ll c$, which is valid for galaxies. Here, we assign each galaxy a single LoS velocity, neglecting internal motions such as rotation.
 
\subsection{Kinematics of G0–G3}\label{sec:kinematics}
There's a redshift difference $\Delta z\sim 0.001-0.003$ among the galaxy members in CGG-z7. If it's due to $z_{\mathrm{cos}}$, their distance along LoS would be greater than $\sim 90$ kpc, which is substantially larger than their projected distance ($<8$ kpc). Such a configuration is unlikely to appear in a gravitationally bound system. Instead, our analysis favors a kinematic origin for the redshift differences. Assuming that all members of CGG-z7 share the same cosmological redshift $z_{\mathrm{cos}}$ recognizing that a self-gravitating galaxy group should have a net LoS velocity of zero, we estimate $z_{\rm cos}  $ as a luminosity-weighted average of the spectroscopic redshifts: 
\begin{equation}
    \hat{z}_\mathrm{cos} = \dfrac{\sum_i m_i z_{\mathrm{spec},i}}{\sum_i m_i}\sim\dfrac{\sum_i L_i z_{\mathrm{spec},i}}{\sum_i L_i}
\end{equation}
where the summation is over G0--G3\footnote{This is reasonable, as G4 and G5 are significantly fainter than G0--G3, as indicated by their non-detection in the grism spectra, implying that the group's total mass is dominated by the brighter galaxies. }, and we adopt the F444W flux as a proxy for luminosity $  L_i  $ for each member. We simply assume the mass of each member is proportional to its F444W flux.
Analysis gives $z_{\mathrm{cos}}=7.040\pm0.001$ for the whole group, and $v_i=-103.0,-33.7,+40.9,+78.3~(\pm35.6)$ km/s for G0--G3, respectively. 

If we further assuming that the system is virialized, we calculate a velocity dispersion of $93.7\pm{31.7}$ km/s using gapper algorithm\footnote{Uncertainties here are estimated with the jackknife method.} \citep{1990AJ....100...32B}. Combined with the harmonic mean projected separation $R_\mathrm{h}=2.20~\rm{kpc}$, this yields a virial mass of $M_\mathrm{vir}=3\pi\sigma_\mathrm{v}^2R_\mathrm{h}/G\sim4.2\pm2.7\times10^{10}~ M_\odot$ \citep{1985ApJ...298....8H,2012MNRAS.426..296D,2025A&A...699A.329T}.

\subsection{Stellar Masses of G0--G3}\label{sec:stellar}
Here we derive stellar masses and other properties of the group members by performing spectral energy distribution (SED) fitting with the code \texttt{Bagpipes} \citep{2018MNRAS.480.4379C,2019MNRAS.490..417C}, which is capable of jointly fitting photometry and spectra. To ensure robustness, the analysis is restricted to galaxy G0--G3, as G4 and G5 are only marginally detected in the F444W band and thus lack sufficient signal-to-noise ratio for reliable SED constraints. Furthermore, to mitigate potential flux contamination between the closely projected members, we use the model fluxes derived from the multi-band morphological decomposition described in Sec.~\ref{sec:morphology} rather than the raw aperture photometry. To ensure a conservative error estimate that accounts for background noise not fully captured by the fitting process alone, we adopt an error floor of 0.0033, 0.0006, and 0.0010 $\mu$Jy for the F115W, F200W, and F444W bands, respectively. This floor is determined by calculating the standard deviation of fluxes measured from random apertures of 0.3 kpc radius placed on blank regions of the sky background. The joint fit incorporates the model spectra described in Sec.~\ref{sec:line}, serving as an additional constraint to ensure consistency across photometric and spectroscopic data. 

During the fitting, we assume a delayed-tau star formation history and a Calzetti extinction law \citep{1994ApJ...429..582C}. The model is evaluated over the following grids: dust attenuation $A_V$ from 0 to 2 mag, stellar population age from 1 to 500 Myr, delay timescale from 10 to 300 Myr, metallicity from 0.2 to $2~ Z_\odot$, ionization parameter $\log U$ from $-3$ to $-1$. The redshift for each source is fixed to $z_{\mathrm{cos}}$ determined in Sec.~\ref{sec:kinematics}. The results of the SED fitting, including the posterior median and credible intervals for key parameters, are summarized in Tab.~\ref{tab:SED}.

The stellar masses are reasonably well constrained, with $\log(M_*/M_\odot)$ ranging from $\sim9.0~\text{to}~9.4$. All galaxies show significant star formation, with SFRs between $\sim5$ and 14 $M_\odot\text{yr}^{-1}$. A notable result is the uniformly low stellar metallicity $<0.2~Z_\odot$ and a high ionization parameter $\log U>-2$ across the group, which is consistent with the extreme emission-line ratios described in Sec.~\ref{sec:line}. We note, however, that the SED properties of G1—including its stellar mass—may be influenced by an AGN. The posterior distributions for age remain broad and prior-sensitive. The results may suffer from degeneracies among age, dust, and star formation history in such data-sparse SED modeling. Deeper, multi-band observations with higher SNR are required to break these degeneracies and to obtain more robust constraints on the stellar populations and dust properties of these galaxies.

The SED fitting results for G0–G3 yield a total stellar mass for the group of $\log(M_*/M_\odot)=9.80^{+0.19}_{-0.21}$. Combined with the total virial mass estimate of $\sim4.2\pm2.7\times10^{10}~ M_\odot$ from Sec.~\ref{sec:kinematics}, we derive a stellar-to-dynamical mass fraction of $M_*/M_\mathrm{vir}\sim 0.15^{+0.20}_{-0.08}$.

This value is extraordinarily high, exceeding the typical range of 0.01 to 0.05 for galaxies and clusters\footnote{We note the potential risk of overestimating $M_*$ of G1. But even without G1, the total stellar mass of the group is $\log(M_*/M_\odot)=9.59^{+0.19}_{-0.20}$, yielding $M_*/M_\mathrm{vir}\sim 0.09^{+0.13}_{-0.05}$.} in the local and high-redshift universe \citep{2020A&A...634A.135G,2025A&A...702A.163P}.
The most compelling interpretation is that the classical virial mass estimator is severely biased low because the system is not in dynamical equilibrium. The extremely high $M_*/M_\mathrm{vir}$ serves as independent, strong evidence that CGG-z7 is a dynamically young system. The low measured velocity dispersion likely reflects non-virial, possibly merging kinematics, rather than a relaxed gravitational potential.  It is consistent with a scenario where the group is undergoing a major merger.

\begin{table*}[htbp]
    \centering
    \caption{SED fitting results of G0--G3 from \texttt{Bagpipes}.}
    \label{tab:SED}
    \begin{tabular}{lcccccc}
    \toprule
    Galaxy & SFR[$M_\odot \text{yr}^{-1}$]& $\log(M_*/M_\odot)$ & $A_V$ & Age[Gyr]& $Z [Z_\odot]$ & $\log U$ \\
    \midrule
    G0 & $6.88^{+3.93}_{-2.32}$ & $9.08^{+0.19}_{-0.19}$ & $0.87^{+0.36}_{-0.25}$ & $0.28^{+0.12}_{-0.11}$ & $0.13^{+0.04}_{-0.05}$ & $-1.86^{+0.55}_{-0.61}$ \\
    G1 & $14.94^{+3.68}_{-4.15}$ & $9.38^{+0.17}_{-0.23}$ & $1.74^{+0.17}_{-0.26}$ & $0.26^{+0.13}_{-0.12}$ & $0.19^{+0.01}_{-0.02}$ & $-1.37^{+0.27}_{-0.46}$ \\
    G2 & $14.20^{+5.58}_{-4.33}$ & $9.29^{+0.20}_{-0.21}$ & $1.61^{+0.23}_{-0.30}$ & $0.22^{+0.14}_{-0.11}$ & $0.15^{+0.03}_{-0.03}$ & $-1.76^{+0.53}_{-0.55}$ \\
    G3 & $5.01^{+2.05}_{-1.39}$ & $8.89^{+0.18}_{-0.21}$ & $0.32^{+0.24}_{-0.20}$ & $0.24^{+0.12}_{-0.10}$ & $0.17^{+0.02}_{-0.03}$ & $-1.68^{+0.50}_{-0.46}$ \\

    \bottomrule
    \end{tabular}
\end{table*}

\subsection{Emission-Line Properties}\label{sec:line}
To obtain robust emission-line fluxes for individual member galaxies while mitigating potential contamination from their closely projected neighbours, we perform two-dimensional spectral fitting again using \texttt{pysersic}. We focus on the H$\beta$ and [O~\textsc{iii}]$\lambda\lambda4959,5007$ features, with cutouts of the 2D spectra for galaxies G0–G3 presented in Fig.~\ref{fig:spec}. Given its highest signal-to-noise ratio (S/N), the [O~\textsc{iii}]$\lambda5007$ line is used as a reference to define the fitting priors, including the spatial position of each galaxy's emission. Here we set all the morphology parameters \texttt{xc,yc,theta,n,ellip,r\_eff} of an individual galaxy to be the same for different lines, allowing only the line fluxes to vary freely. 

\begin{table*}[htbp]
    \centering
    \caption{Line Flux (unit: $10^{-18}~ \rm{erg~s^{-1}~cm^{-2}}$). Here R3 is defined as $($[O{\sc iii}]$\lambda$4959+[O{\sc iii}]$\lambda$5007$)/$H$\beta$.}
    \label{tab:group_flux}
    \begin{tabular}{lccccc}
        \toprule
        Galaxy & H$\beta$ & [O {\sc iii}]$\lambda$4959 & [O {\sc iii}]$\lambda$5007 & R3 \\
        \midrule
        G0 & $0.89^{+0.20}_{-0.17}$ & $1.35^{+0.21}_{-0.23}$ & $3.46^{+0.26}_{-0.24}$ & $5.38^{+1.49}_{-0.98}$  \\
        G1 & $0.46^{+0.14}_{-0.15}$ & $2.01^{+0.25}_{-0.22}$ & $6.18^{+0.43}_{-0.38}$ & $17.85^{+8.22}_{-4.40}$ \\
        G2 & $1.11^{+0.20}_{-0.21}$ & $1.36^{+0.21}_{-0.23}$ & $5.29^{+0.37}_{-0.33}$ & $5.98^{+1.41}_{-0.98}$ \\
        G3 & $0.99^{+0.21}_{-0.19}$ & $1.37^{+0.21}_{-0.20}$ & $5.91^{+0.38}_{-0.26}$ & $7.38^{+1.95}_{-1.35}$ \\
        
        \bottomrule
    \end{tabular}
\end{table*}

The resulting line fluxes are summarized in Tab.~\ref{tab:group_flux}. As a key quality check, the measured flux ratios [O~\textsc{iii}]$\lambda4959$ / [O~\textsc{iii}]$\lambda5007$ for the four galaxies are consistent with the theoretical value of 1:2.98 within $2\sigma$ (1.08$\sigma$, 0.09$\sigma$, 1.29$\sigma$, and 1.84$\sigma$ for G0–G3, respectively). A comparison of data and model is shown in Fig.~\ref{fig:line_model}.
From these fluxes, we calculate the line ratio R3 $=($[O{\sc iii}]$\lambda4959$ + [O{\sc iii}]$\lambda5007)/$H$\beta$\footnote{We do not apply a dust attenuation correction here.}, finding values in the range of approximately 5 to 18.

The elevated R3 ratios are generally indicative of extreme conditions in the ionized gas. They commonly signal a hard radiation field with high ionization parameters and/or exceptionally low gas-phase metallicity \citep{2024ApJ...975..245C,2025NatAs...9..155Z,2025arXiv251013952M}. While R3 is commonly used as a gas-phase metallicity indicator, its application here is problematic for several reasons. Most established empirical calibrations are described with a double-branched relation with a second or third-order polynomial function peaking at R3 $\sim 8.4$ \citep[e.g.][]{2023MNRAS.526.3504H,2024ApJ...962...24S,2025ApJ...985...24C}\footnote{Some of them define R3 as [O{\sc iii}]$\lambda5007/$H$\beta$, which is 3/4 of our R3.}. As our R3 falls outside the valid domain or near the turnover point of these functions, and because the relation is flat near the peak, any attempt to invert it would yield degenerate solutions with large uncertainties. Moreover, such calibrations often exhibit significant intrinsic scatter, particularly at high redshifts. Low S/N of H$\beta$ ($\lesssim1.5$) and [O~\textsc{iii}] $\lambda4959$ ($\sim2$) also blur the measurements. So we refrain from deriving quantitative metallicity estimates for individual galaxies in this work. The high R3 values are reported here primarily to characterize the extreme ionized gas conditions and low-metallicity within CGG-z7.

The case of G1 warrants specific attention. It exhibits the most extreme line ratio in the group (R3 $\sim17.85$), a notably high value compared to typical star-forming galaxies \citep{2014MNRAS.444.3466S,2023MNRAS.526.3504H,2023ApJS..269...33N,2024ApJ...962...24S,2025ApJ...985...24C,2025ApJS..280...62L}. This, combined with its status as the most compact and barely resolved source (see Sec.~\ref{sec:morphology}) and its high ionization parameter ($\log U\sim -1.37$ from SED fitting, albeit with large uncertainty), makes G1 a compelling candidate for hosting an obscured AGN.

\section{Summary and Discussion}\label{sec:summary}
In this paper, we report the discovery of CGG-z7, a compact galaxy group at redshift $\sim7.04$, identified in public JWST/NIRCam imaging and grism data from the POPPIES program (PID 5398). The group comprises at least six member candidates detected in deep F200W imaging, four of which (G0–G3) are spectroscopically confirmed via their H$\beta$ and [O{\sc iii}] doublet emission in the F444W grism spectra. With a projected size of only $\sim 7.8\times5.7$ kpc$^2$, CGG-z7 is among the most compact and highest-redshift galaxy groups known to date. 

\subsection{Properties of CGG-z7}
Multi-band morphological analysis with \texttt{pysersic} reveals that the group members have Sérsic indices in the range $n\sim0.8 - 3.4$, all with compact sizes that $r_\mathrm{eff}\lesssim 0.8$ kpc. Spectral energy distribution (SED) fitting with \texttt{Bagpipes} indicates that all G0--G3 are star-forming, metal-poor, and exhibit high ionization parameters $\log U> -2$. These properties are corroborated by strong [O {\sc iii}] emission, with line ratios R3 $\sim 5-18$ suggesting extreme ionization conditions and low gas-phase metallicity in the ionized gas.

The group’s systemic cosmological redshift is precisely measured as $z_{\mathrm{cos}}=7.040\pm0.001$. From the velocity differences of the members, we derive a line-of-sight velocity dispersion of $93.7\pm{31.7}$ km/s. Under the assumption of virial equilibrium, this implies a dynamical mass of $M_\mathrm{vir}\approx 4.2\pm2.7\times10^{10}~ M_\odot$. However, the resulting stellar-to-dynamical mass ratio, $M_*/M_\mathrm{vir}\sim0.15^{+0.20}_{-0.08}$, is three times higher than in typical relaxed systems. This large discrepancy indicates that the classical virial mass estimator is likely biased low because CGG-z7 is not in dynamical equilibrium, pointing instead to a young, unrelaxed system that may be undergoing a merger.

\subsection{G1 Hosts an AGN}
In particular, the extreme properties of member G1—including its compact morphology, exceptionally high [O {\sc iii}]/H$\beta$ ratio and ionization parameter—strongly suggest it hosts an AGN instead of pure star formation, consistent with high-redshift AGN diagnostics in dense environments that $\log($[O {\sc iii}]/H$\beta) \gtrsim 0.8-1.5$ and $\log U\gtrsim-2$ \citep{2024ApJ...975..245C,2025A&A...700A..12M,2025A&A...697A.175S}. This finding aligns with the established trend that galaxy interactions in compact groups substantially enhance both AGN activity and star formation, making the emergence of such an obscured AGN in CGG-z7 not uncommon \citep[e.g.][]{2010AJ....139.1199M,2025arXiv251014743P,2025A&A...704A.101S,2026ApJ...997..208S}. 
Probably fueled by frequent mergers and gas inflows, the AGN is expected to drive powerful feedback processes, suppressing future star formation and accelerating the depletion of the interstellar medium.

\subsection{Fate of the Compact Galaxy Group}
The properties of CGG-z7—its high star formation rate, uniformly extreme ionization state, and likely merging state—suggest it may be a direct progenitor of the massive quiescent or post-starburst galaxies observed at $z\sim 5-6$ \citep{2022A&A...665L...7S,2024A&A...683L...4J,2025arXiv251113650S}. Within the deep potential well, 
the intense radiation field, alongside the merger or interaction triggered starburst and AGN feedback, could rapidly consume or expel the remaining gas
\citep{2021MNRAS.504.4533W,2022MNRAS.513.3252A,2024ApJ...963...49K,2024A&A...683L...4J,2024A&A...690A..55S}. Subsequent dynamical relaxation and quenching, possibly accelerated by the merger itself, would transform this compact group into a single, massive spheroidal quenched galaxy, i.e. a ``Red Nugget", on a timescale of a few hundred million years \citep{2023A&A...670L..11J,2025A&A...694A.218B,2025arXiv251113650S}.

CGG-z7 establishes a pivotal high-redshift benchmark; future deep, multi-wavelength interal field spectrograph (IFU) is  needed to unveil its full diagnostic potential for early galaxy formation and evolution.

\begin{acknowledgments}
This work is supported by Z.C. funded by National Key R\&D Program of China, grantnumber 2023YFA1605600, National Natural Science Foundation of China, grantnumber 12525303, and Tsinghua University Initiative Scientific ResearchProgram. J.W. was funded by National Natural Science Foundation of China, grant number 62222508 and 62525506. Z.C. and J.W. were funded by New Cornerstone Science Foundation through the XPLORER PRIZE.

This work is based in part on observations made with the NASA/ESA/CSA James Webb Space Telescope. The data were obtained from the Mikulski Archive for Space Telescopes at the Space Telescope Science Institute. The observations are associated with program \#5398. The authors acknowledge the POPPIES team for developing their observing program with a zero-exclusive-access period.

\end{acknowledgments}





%
\facilities{JWST (NIRCam)}


\software{astropy \citep{2013A&A...558A..33A,2018AJ....156..123A,2022ApJ...935..167A},  
          pysersic \citep{2023JOSS....8.5703P},
          photutils \citep{larry_bradley_2025_14889440},
          jwst \citep{Bushouse2025},
          Bagpipes \citep{2018MNRAS.480.4379C,2019MNRAS.490..417C},
          stpsf \citep{Perrin2025},
          numpy \citep{harris2020array},
          scipy \citep{2020SciPy-NMeth},
          MyFilter \citep{li_2023_10210202}.
          }


\appendix
\section{Supplementary Figures}
\begin{figure*}[htbp]
    \centering
    \includegraphics[width=0.96\linewidth]{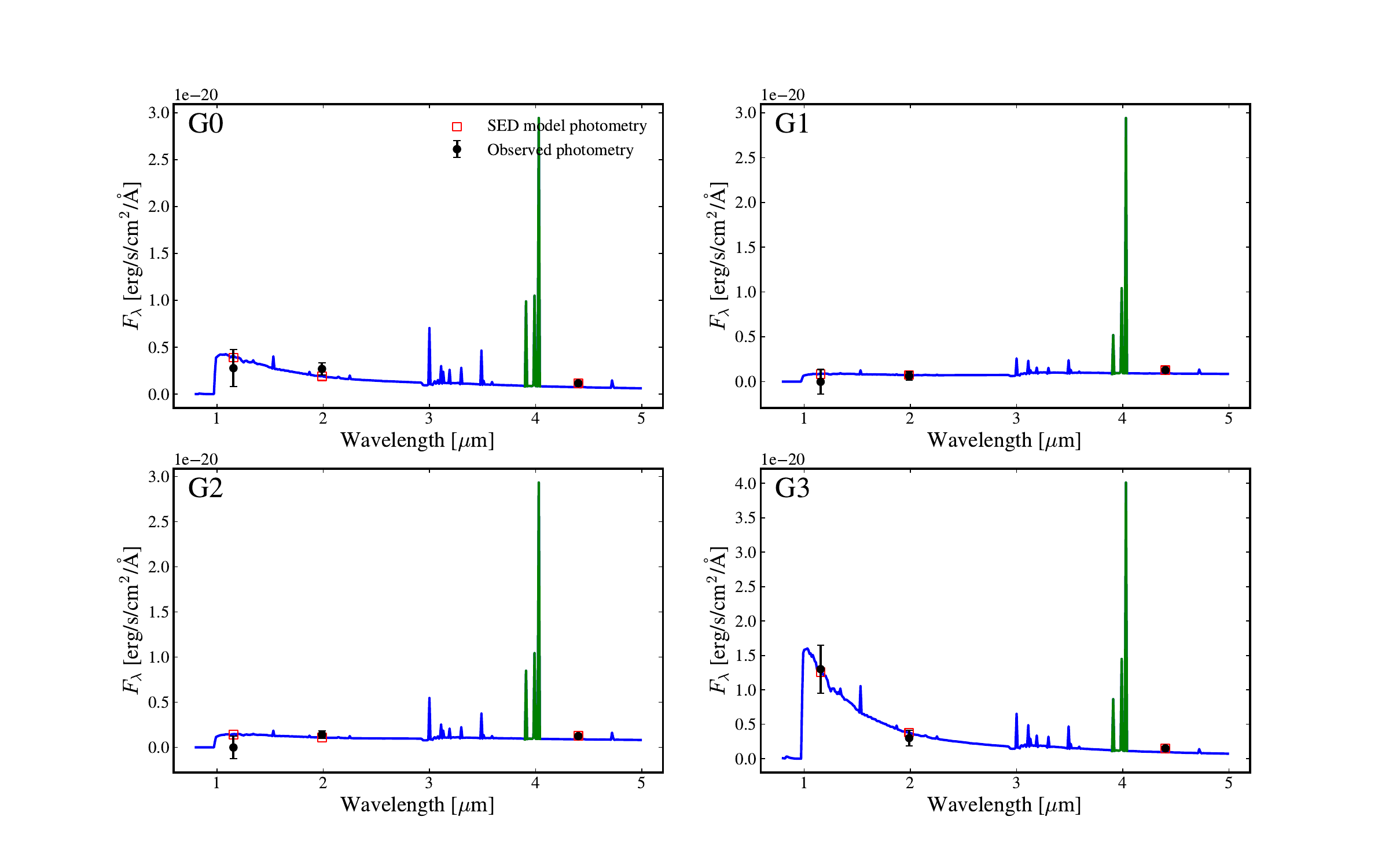}
    \caption{SEDs of G0--G3 in CGG-z7. The observed photometry means the model fluxes derived by \texttt{pysersic}. The best-fit models (blue lines) are obtained by jointly fitting the photometry and the F444W grism spectra; the segments highlighted in green indicate the wavelength ranges constrained by the grism data.}
    \label{fig:sed}
\end{figure*}

\begin{figure*}[htbp]
    \centering
    \includegraphics[width=0.8\linewidth]{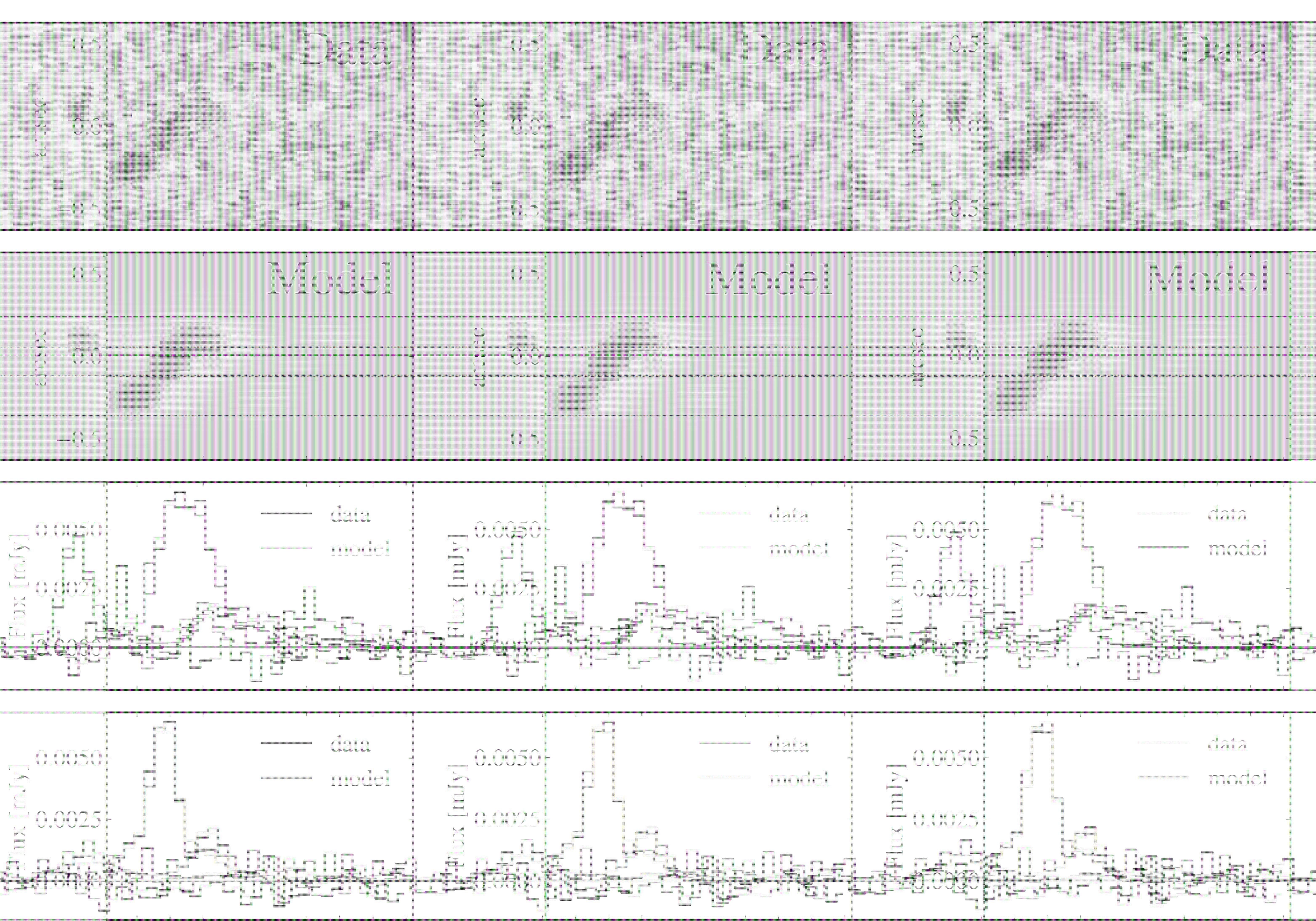}
    \caption{F444W grism spectra (black) and the corresponding model spectra from \texttt{pysersic} (colored lines). The green, red, and blue lines show the 1D model spectra for the regions between the pairs of horizontal dashed lines of the same color.}
    \label{fig:line_model}
\end{figure*}

Table~\ref{tab:morphology_fit} presents the full morphological fitting results for all six candidate members of CGG-z7, complementing the summary in Section~\ref{sec:morphology}. 
Figure~\ref{fig:sed} shows the full SED fitting results for G0--G3, and Figure~\ref{fig:line_model} compares the grism spectra and the corresponding models.

\begin{table*}[htbp]
    \centering
    \caption{Morphological parameters of the group members by \texttt{pysersic}.}
    \label{tab:morphology_fit}
    \begin{tabular}{lcccccc}
    \toprule
    Source & Band & Flux [nJy] & $n$ & $r_{\rm eff}$ [kpc] \\
    \midrule
    \multirow{3}{*}{G0} & F444W & $75.28^{+4.96}_{-4.25}$ & $1.25^{+0.42}_{-0.24}$ & $0.43^{+0.05}_{-0.04}$ \\
    & F200W & $35.45^{+3.18}_{-2.81}$ & $1.55^{+0.60}_{-0.38}$ & $0.31^{+0.04}_{-0.04}$ \\
    & F115W & $12.31^{+8.70}_{-8.27}$ & $1.60^{+0.78}_{-0.47}$ & $0.30^{+0.07}_{-0.07}$ \\
    \addlinespace
    
    \multirow{3}{*}{G1} & F444W & $83.13^{+6.12}_{-5.30}$ & $2.42^{+1.35}_{-0.82}$ & $0.29^{+0.05}_{-0.05}$ \\
    & F200W & $8.28^{+1.92}_{-1.95}$ & $3.26^{+1.42}_{-1.20}$ & $0.24^{+0.12}_{-0.09}$ \\
    & F115W & $-0.10^{+0.68}_{-0.64}$ & $3.42^{+1.74}_{-1.38}$ & $0.23^{+0.15}_{-0.13}$ \\
    \addlinespace
    \multirow{3}{*}{G2} & F444W & $81.07^{+5.36}_{-5.53}$ & $0.97^{+0.22}_{-0.15}$ & $0.47^{+0.04}_{-0.04}$ \\
    & F200W & $18.73^{+2.70}_{-3.02}$ & $1.41^{+0.59}_{-0.36}$ & $0.46^{+0.09}_{-0.08}$ \\
    & F115W & $-0.04^{+0.63}_{-0.66}$ & $1.49^{+0.83}_{-0.43}$ & $0.46^{+0.12}_{-0.12}$ \\
    \addlinespace
    \multirow{3}{*}{G3} & F444W & $97.53^{+4.87}_{-4.95}$ & $0.84^{+0.13}_{-0.08}$ & $0.58^{+0.04}_{-0.04}$ \\
    & F200W & $39.57^{+4.14}_{-3.71}$ & $1.03^{+0.25}_{-0.14}$ & $0.79^{+0.09}_{-0.08}$ \\
    & F115W & $57.78^{+14.78}_{-15.38}$ & $1.05^{+0.31}_{-0.18}$ & $0.90^{+0.12}_{-0.13}$ \\
    \addlinespace
    \multirow{3}{*}{G4} & F444W & $33.72^{+5.60}_{-5.41}$ & $1.07^{+0.41}_{-0.22}$ & $0.78^{+0.13}_{-0.13}$ \\
    & F200W & $37.29^{+5.40}_{-4.81}$ & $1.38^{+0.42}_{-0.29}$ & $0.77^{+0.11}_{-0.12}$ \\
    & F115W & $-0.03^{+0.67}_{-0.64}$ & $1.45^{+0.58}_{-0.37}$ & $0.77^{+0.14}_{-0.15}$ \\
    \addlinespace
    \multirow{3}{*}{G5} & F444W & $14.58^{+3.59}_{-2.87}$ & $3.16^{+2.11}_{-1.44}$ & $0.22^{+0.15}_{-0.15}$ \\
    & F200W & $12.24^{+2.61}_{-2.54}$ & $3.12^{+1.87}_{-1.25}$ & $0.36^{+0.11}_{-0.10}$ \\
    & F115W & $-0.05^{+0.69}_{-0.69}$ & $3.14^{+1.90}_{-1.34}$ & $0.39^{+0.14}_{-0.12}$ \\
    \bottomrule
    
    \end{tabular}
\end{table*}

\bibliography{sample701}{}

@ARTICLE{2022ApJ...935..167A,
       author = {{Astropy Collaboration} and {Price-Whelan}, Adrian M. and {Lim}, Pey Lian and {Earl}, Nicholas and {Starkman}, Nathaniel and {Bradley}, Larry and {Shupe}, David L. and {Patil}, Aarya A. and {Corrales}, Lia and {Brasseur}, C.~E. and {N{\"o}the}, Maximilian and {Donath}, Axel and {Tollerud}, Erik and {Morris}, Brett M. and {Ginsburg}, Adam and {Vaher}, Eero and {Weaver}, Benjamin A. and {Tocknell}, James and {Jamieson}, William and {van Kerkwijk}, Marten H. and {Robitaille}, Thomas P. and {Merry}, Bruce and {Bachetti}, Matteo and {G{\"u}nther}, H. Moritz and {Aldcroft}, Thomas L. and {Alvarado-Montes}, Jaime A. and {Archibald}, Anne M. and {B{\'o}di}, Attila and {Bapat}, Shreyas and {Barentsen}, Geert and {Baz{\'a}n}, Juanjo and {Biswas}, Manish and {Boquien}, M{\'e}d{\'e}ric and {Burke}, D.~J. and {Cara}, Daria and {Cara}, Mihai and {Conroy}, Kyle E. and {Conseil}, Simon and {Craig}, Matthew W. and {Cross}, Robert M. and {Cruz}, Kelle L. and {D'Eugenio}, Francesco and {Dencheva}, Nadia and {Devillepoix}, Hadrien A.~R. and {Dietrich}, J{\"o}rg P. and {Eigenbrot}, Arthur Davis and {Erben}, Thomas and {Ferreira}, Leonardo and {Foreman-Mackey}, Daniel and {Fox}, Ryan and {Freij}, Nabil and {Garg}, Suyog and {Geda}, Robel and {Glattly}, Lauren and {Gondhalekar}, Yash and {Gordon}, Karl D. and {Grant}, David and {Greenfield}, Perry and {Groener}, Austen M. and {Guest}, Steve and {Gurovich}, Sebastian and {Handberg}, Rasmus and {Hart}, Akeem and {Hatfield-Dodds}, Zac and {Homeier}, Derek and {Hosseinzadeh}, Griffin and {Jenness}, Tim and {Jones}, Craig K. and {Joseph}, Prajwel and {Kalmbach}, J. Bryce and {Karamehmetoglu}, Emir and {Ka{\l}uszy{\'n}ski}, Miko{\l}aj and {Kelley}, Michael S.~P. and {Kern}, Nicholas and {Kerzendorf}, Wolfgang E. and {Koch}, Eric W. and {Kulumani}, Shankar and {Lee}, Antony and {Ly}, Chun and {Ma}, Zhiyuan and {MacBride}, Conor and {Maljaars}, Jakob M. and {Muna}, Demitri and {Murphy}, N.~A. and {Norman}, Henrik and {O'Steen}, Richard and {Oman}, Kyle A. and {Pacifici}, Camilla and {Pascual}, Sergio and {Pascual-Granado}, J. and {Patil}, Rohit R. and {Perren}, Gabriel I. and {Pickering}, Timothy E. and {Rastogi}, Tanuj and {Roulston}, Benjamin R. and {Ryan}, Daniel F. and {Rykoff}, Eli S. and {Sabater}, Jose and {Sakurikar}, Parikshit and {Salgado}, Jes{\'u}s and {Sanghi}, Aniket and {Saunders}, Nicholas and {Savchenko}, Volodymyr and {Schwardt}, Ludwig and {Seifert-Eckert}, Michael and {Shih}, Albert Y. and {Jain}, Anany Shrey and {Shukla}, Gyanendra and {Sick}, Jonathan and {Simpson}, Chris and {Singanamalla}, Sudheesh and {Singer}, Leo P. and {Singhal}, Jaladh and {Sinha}, Manodeep and {Sip{\H{o}}cz}, Brigitta M. and {Spitler}, Lee R. and {Stansby}, David and {Streicher}, Ole and {{\v{S}}umak}, Jani and {Swinbank}, John D. and {Taranu}, Dan S. and {Tewary}, Nikita and {Tremblay}, Grant R. and {de Val-Borro}, Miguel and {Van Kooten}, Samuel J. and {Vasovi{\'c}}, Zlatan and {Verma}, Shresth and {de Miranda Cardoso}, Jos{\'e} Vin{\'\i}cius and {Williams}, Peter K.~G. and {Wilson}, Tom J. and {Winkel}, Benjamin and {Wood-Vasey}, W.~M. and {Xue}, Rui and {Yoachim}, Peter and {Zhang}, Chen and {Zonca}, Andrea and {Astropy Project Contributors}},
        title = "{The Astropy Project: Sustaining and Growing a Community-oriented Open-source Project and the Latest Major Release (v5.0) of the Core Package}",
      journal = {\apj},
         year = 2022,
        month = aug,
       volume = {935},
       number = {2},
          eid = {167},
        pages = {167},
          doi = {10.3847/1538-4357/ac7c74},
archivePrefix = {arXiv},
       eprint = {2206.14220},
 primaryClass = {astro-ph.IM},
       adsurl = {https://ui.adsabs.harvard.edu/abs/2022ApJ...935..167A}
}

@ARTICLE{2018AJ....156..123A,
       author = {{Astropy Collaboration} and {Price-Whelan}, A.~M. and {Sip{\H{o}}cz}, B.~M. and {G{\"u}nther}, H.~M. and {Lim}, P.~L. and {Crawford}, S.~M. and {Conseil}, S. and {Shupe}, D.~L. and {Craig}, M.~W. and {Dencheva}, N. and {Ginsburg}, A. and {VanderPlas}, J.~T. and {Bradley}, L.~D. and {P{\'e}rez-Su{\'a}rez}, D. and {de Val-Borro}, M. and {Aldcroft}, T.~L. and {Cruz}, K.~L. and {Robitaille}, T.~P. and {Tollerud}, E.~J. and {Ardelean}, C. and {Babej}, T. and {Bach}, Y.~P. and {Bachetti}, M. and {Bakanov}, A.~V. and {Bamford}, S.~P. and {Barentsen}, G. and {Barmby}, P. and {Baumbach}, A. and {Berry}, K.~L. and {Biscani}, F. and {Boquien}, M. and {Bostroem}, K.~A. and {Bouma}, L.~G. and {Brammer}, G.~B. and {Bray}, E.~M. and {Breytenbach}, H. and {Buddelmeijer}, H. and {Burke}, D.~J. and {Calderone}, G. and {Cano Rodr{\'\i}guez}, J.~L. and {Cara}, M. and {Cardoso}, J.~V.~M. and {Cheedella}, S. and {Copin}, Y. and {Corrales}, L. and {Crichton}, D. and {D'Avella}, D. and {Deil}, C. and {Depagne}, {\'E}. and {Dietrich}, J.~P. and {Donath}, A. and {Droettboom}, M. and {Earl}, N. and {Erben}, T. and {Fabbro}, S. and {Ferreira}, L.~A. and {Finethy}, T. and {Fox}, R.~T. and {Garrison}, L.~H. and {Gibbons}, S.~L.~J. and {Goldstein}, D.~A. and {Gommers}, R. and {Greco}, J.~P. and {Greenfield}, P. and {Groener}, A.~M. and {Grollier}, F. and {Hagen}, A. and {Hirst}, P. and {Homeier}, D. and {Horton}, A.~J. and {Hosseinzadeh}, G. and {Hu}, L. and {Hunkeler}, J.~S. and {Ivezi{\'c}}, {\v{Z}}. and {Jain}, A. and {Jenness}, T. and {Kanarek}, G. and {Kendrew}, S. and {Kern}, N.~S. and {Kerzendorf}, W.~E. and {Khvalko}, A. and {King}, J. and {Kirkby}, D. and {Kulkarni}, A.~M. and {Kumar}, A. and {Lee}, A. and {Lenz}, D. and {Littlefair}, S.~P. and {Ma}, Z. and {Macleod}, D.~M. and {Mastropietro}, M. and {McCully}, C. and {Montagnac}, S. and {Morris}, B.~M. and {Mueller}, M. and {Mumford}, S.~J. and {Muna}, D. and {Murphy}, N.~A. and {Nelson}, S. and {Nguyen}, G.~H. and {Ninan}, J.~P. and {N{\"o}the}, M. and {Ogaz}, S. and {Oh}, S. and {Parejko}, J.~K. and {Parley}, N. and {Pascual}, S. and {Patil}, R. and {Patil}, A.~A. and {Plunkett}, A.~L. and {Prochaska}, J.~X. and {Rastogi}, T. and {Reddy Janga}, V. and {Sabater}, J. and {Sakurikar}, P. and {Seifert}, M. and {Sherbert}, L.~E. and {Sherwood-Taylor}, H. and {Shih}, A.~Y. and {Sick}, J. and {Silbiger}, M.~T. and {Singanamalla}, S. and {Singer}, L.~P. and {Sladen}, P.~H. and {Sooley}, K.~A. and {Sornarajah}, S. and {Streicher}, O. and {Teuben}, P. and {Thomas}, S.~W. and {Tremblay}, G.~R. and {Turner}, J.~E.~H. and {Terr{\'o}n}, V. and {van Kerkwijk}, M.~H. and {de la Vega}, A. and {Watkins}, L.~L. and {Weaver}, B.~A. and {Whitmore}, J.~B. and {Woillez}, J. and {Zabalza}, V. and {Astropy Contributors}},
        title = "{The Astropy Project: Building an Open-science Project and Status of the v2.0 Core Package}",
      journal = {\aj},
         year = 2018,
        month = sep,
       volume = {156},
       number = {3},
          eid = {123},
        pages = {123},
          doi = {10.3847/1538-3881/aabc4f},
archivePrefix = {arXiv},
       eprint = {1801.02634},
 primaryClass = {astro-ph.IM},
       adsurl = {https://ui.adsabs.harvard.edu/abs/2018AJ....156..123A}
}

@ARTICLE{2013A&A...558A..33A,
       author = {{Astropy Collaboration} and {Robitaille}, Thomas P. and
         {Tollerud}, Erik J. and {Greenfield}, Perry and {Droettboom}, Michael and
         {Bray}, Erik and {Aldcroft}, Tom and {Davis}, Matt and
         {Ginsburg}, Adam and {Price-Whelan}, Adrian M. and
         {Kerzendorf}, Wolfgang E. and {Conley}, Alexander and {Crighton}, Neil and
         {Barbary}, Kyle and {Muna}, Demitri and {Ferguson}, Henry and
         {Grollier}, Fr{\'e}d{\'e}ric and {Parikh}, Madhura M. and
         {Nair}, Prasanth H. and {Unther}, Hans M. and {Deil}, Christoph and
         {Woillez}, Julien and {Conseil}, Simon and {Kramer}, Roban and
         {Turner}, James E.~H. and {Singer}, Leo and {Fox}, Ryan and
         {Weaver}, Benjamin A. and {Zabalza}, Victor and {Edwards}, Zachary I. and
         {Azalee Bostroem}, K. and {Burke}, D.~J. and {Casey}, Andrew R. and
         {Crawford}, Steven M. and {Dencheva}, Nadia and {Ely}, Justin and
         {Jenness}, Tim and {Labrie}, Kathleen and {Lim}, Pey Lian and
         {Pierfederici}, Francesco and {Pontzen}, Andrew and {Ptak}, Andy and
         {Refsdal}, Brian and {Servillat}, Mathieu and {Streicher}, Ole},
        title = "{Astropy: A community Python package for astronomy}",
      journal = {\aap},
         year = "2013",
        month = "Oct",
       volume = {558},
          eid = {A33},
        pages = {A33},
          doi = {10.1051/0004-6361/201322068},
archivePrefix = {arXiv},
       eprint = {1307.6212},
 primaryClass = {astro-ph.IM},
       adsurl = {https://ui.adsabs.harvard.edu/abs/2013A&A...558A..33A}
}

@ARTICLE{2004ApJ...600L..93G,
       author = {{Giavalisco}, M. and {Ferguson}, H.~C. and {Koekemoer}, A.~M. and {Dickinson}, M. and {Alexander}, D.~M. and {Bauer}, F.~E. and {Bergeron}, J. and {Biagetti}, C. and {Brandt}, W.~N. and {Casertano}, S. and {Cesarsky}, C. and {Chatzichristou}, E. and {Conselice}, C. and {Cristiani}, S. and {Da Costa}, L. and {Dahlen}, T. and {de Mello}, D. and {Eisenhardt}, P. and {Erben}, T. and {Fall}, S.~M. and {Fassnacht}, C. and {Fosbury}, R. and {Fruchter}, A. and {Gardner}, J.~P. and {Grogin}, N. and {Hook}, R.~N. and {Hornschemeier}, A.~E. and {Idzi}, R. and {Jogee}, S. and {Kretchmer}, C. and {Laidler}, V. and {Lee}, K.~S. and {Livio}, M. and {Lucas}, R. and {Madau}, P. and {Mobasher}, B. and {Moustakas}, L.~A. and {Nonino}, M. and {Padovani}, P. and {Papovich}, C. and {Park}, Y. and {Ravindranath}, S. and {Renzini}, A. and {Richardson}, M. and {Riess}, A. and {Rosati}, P. and {Schirmer}, M. and {Schreier}, E. and {Somerville}, R.~S. and {Spinrad}, H. and {Stern}, D. and {Stiavelli}, M. and {Strolger}, L. and {Urry}, C.~M. and {Vandame}, B. and {Williams}, R. and {Wolf}, C.},
        title = "{The Great Observatories Origins Deep Survey: Initial Results from Optical and Near-Infrared Imaging}",
      journal = {\apjl},
         year = 2004,
        month = jan,
       volume = {600},
       number = {2},
        pages = {L93-L98},
          doi = {10.1086/379232},
archivePrefix = {arXiv},
       eprint = {astro-ph/0309105},
 primaryClass = {astro-ph},
       adsurl = {https://ui.adsabs.harvard.edu/abs/2004ApJ...600L..93G}
}

@misc{Bushouse2025,
  author    = {Bushouse, H. and Eisenhamer, J. and Dencheva, N. and Davies, J. and Greenfield, P. and Morrison, J. and Hodge, P. and Simon, B. and Grumm, D. and Droettboom, M. and Slavich, E. and Sosey, M. and Pauly, T. and Miller, T. and Jedrzejewski, R. and Hack, W. and Davis, D. and Crawford, S. and Law, D. and Gordon, K.},
  title     = {JWST Calibration Pipeline},
  year      = {2025},
  howpublished = {Zenodo (CERN European Organization for Nuclear Research)},
  doi       = {10.5281/zenodo.15632984},
  url       = {https://doi.org/10.5281/zenodo.15632984}
}

@software{larry_bradley_2025_14889440,
  author       = {Larry Bradley and
                  Brigitta Sip{\H o}cz and
                  Thomas Robitaille and
                  Erik Tollerud and
                  Z\`e Vin{\'{\i}}cius and
                  Christoph Deil and
                  Kyle Barbary and
                  Tom J Wilson and
                  Ivo Busko and
                  Axel Donath and
                  Hans Moritz G{\"u}nther and
                  Mihai Cara and
                  P. L. Lim and
                  Sebastian Me{\ss}linger and
                  Zach Burnett and
                  Simon Conseil and
                  Michael Droettboom and
                  Azalee Bostroem and
                  E. M. Bray and
                  Lars Andersen Bratholm and
                  William Jamieson and
                  Adam Ginsburg and
                  Geert Barentsen and
                  Matt Craig and
                  Sergio Pascual and
                  Shivangee Rathi and
                  Marshall Perrin and
                  Brett M. Morris},
  title        = {astropy/photutils: 2.2.0},
  month        = feb,
  year         = 2025,
  publisher    = {Zenodo},
  version      = {2.2.0},
  doi          = {10.5281/zenodo.14889440},
  url          = {https://doi.org/10.5281/zenodo.14889440},
  swhid        = {swh:1:dir:11159107f27a28985192ed1118b1f2055709d093
                   ;origin=https://doi.org/10.5281/zenodo.596036;visi
                   t=swh:1:snp:ae8c4a55d349d43e53cfe9ce92e678fcfe840f
                   3b;anchor=swh:1:rel:0117f67e8888adcdfc85308287dd9c
                   854b466389;path=astropy-photutils-ffb96c5
                  },
}

@software{li_2023_10210202,
  author       = {Li, Mingyu},
  title        = {MyFilter: A Web Application for Interactive
                   Visualization of Astronomical Filter Transmission
                   Curves
                  },
  month        = nov,
  year         = 2023,
  publisher    = {Zenodo},
  version      = {1.0.0},
  doi          = {10.5281/zenodo.10210202},
  url          = {https://doi.org/10.5281/zenodo.10210202},
}

@ARTICLE{2023A&A...674A..22G,
       author = {{Gavras}, Panagiotis and {Rimoldini}, Lorenzo and {Nienartowicz}, Krzysztof and {de Fombelle}, Gr{\'e}gory Jevardat and {Holl}, Berry and {{\'A}brah{\'a}m}, P{\'e}ter and {Audard}, Marc and {Carnerero}, Maria I. and {Clementini}, Gisella and {De Ridder}, Joris and {Distefano}, Elisa and {Garcia-Lario}, Pedro and {Garofalo}, Alessia and {K{\'o}sp{\'a}l}, {\'A}gnes and {Kruszy{\'n}ska}, Katarzyna and {Kun}, M{\'a}ria and {Lecoeur-Ta{\"\i}bi}, Isabelle and {Marton}, G{\'a}bor and {Mazeh}, Tsevi and {Mowlavi}, Nami and {Raiteri}, Claudia M. and {Ripepi}, Vincenzo and {Szabados}, L{\'a}szl{\'o} and {Zucker}, Shay and {Eyer}, Laurent},
        title = "{Gaia Data Release 3. Cross-match of Gaia sources with variable objects from the literature}",
      journal = {\aap},
         year = 2023,
        month = jun,
       volume = {674},
          eid = {A22},
        pages = {A22},
          doi = {10.1051/0004-6361/202244367},
archivePrefix = {arXiv},
       eprint = {2207.01946},
 primaryClass = {astro-ph.IM},
       adsurl = {https://ui.adsabs.harvard.edu/abs/2023A&A...674A..22G}
}

@ARTICLE{2023ApJ...953...53S,
       author = {{Sun}, Fengwu and {Egami}, Eiichi and {Pirzkal}, Nor and {Rieke}, Marcia and {Baum}, Stefi and {Boyer}, Martha and {Boyett}, Kristan and {Bunker}, Andrew J. and {Cameron}, Alex J. and {Curti}, Mirko and {Eisenstein}, Daniel J. and {Gennaro}, Mario and {Greene}, Thomas P. and {Jaffe}, Daniel and {Kelly}, Doug and {Koekemoer}, Anton M. and {Kumari}, Nimisha and {Maiolino}, Roberto and {Maseda}, Michael and {Perna}, Michele and {Rest}, Armin and {Robertson}, Brant E. and {Schlawin}, Everett and {Smit}, Renske and {Stansberry}, John and {Sunnquist}, Ben and {Tacchella}, Sandro and {Williams}, Christina C. and {Willmer}, Christopher N.~A.},
        title = "{First Sample of H{\ensuremath{\alpha}}+[O III]{\ensuremath{\lambda}}5007 Line Emitters at z > 6 Through JWST/NIRCam Slitless Spectroscopy: Physical Properties and Line-luminosity Functions}",
      journal = {\apj},
         year = 2023,
        month = aug,
       volume = {953},
       number = {1},
          eid = {53},
        pages = {53},
          doi = {10.3847/1538-4357/acd53c},
archivePrefix = {arXiv},
       eprint = {2209.03374},
 primaryClass = {astro-ph.GA},
       adsurl = {https://ui.adsabs.harvard.edu/abs/2023ApJ...953...53S}
}

@ARTICLE{2025arXiv250315587S,
       author = {{Sun}, Fengwu and {Fudamoto}, Yoshinobu and {Lin}, Xiaojing and {Helton}, Jakob M. and {Hsiao}, Tiger Yu-Yang and {Egami}, Eiichi and {Akhtarkavan}, Arshia and {Bunker}, Andrew J. and {Cai}, Zheng and {DeCoursey}, Christa and {Eisenstein}, Daniel J. and {Fan}, Xiaohui and {Harikane}, Yuichi and {Ji}, Zhiyuan and {Jin}, Xiangyu and {Liu}, Weizhe and {Liu}, Yichen and {Ma}, Zheng and {Maiolino}, Roberto and {Ouchi}, Masami and {Tee}, Wei Leong and {Wang}, Feige and {Willmer}, Christopher N.~A. and {Wu}, Yunjing and {Xu}, Yi and {Yang}, Jinyi and {Zhang}, Junyu and {Zhu}, Yongda},
        title = "{Slitless Areal Pure-Parallel HIgh-Redshift Emission Survey (SAPPHIRES): Early Data Release of Deep JWST/NIRCam Images and Spectra in MACS J0416 Parallel Field}",
      journal = {arXiv e-prints},
         year = 2025,
        month = mar,
          eid = {arXiv:2503.15587},
        pages = {arXiv:2503.15587},
          doi = {10.48550/arXiv.2503.15587},
archivePrefix = {arXiv},
       eprint = {2503.15587},
 primaryClass = {astro-ph.GA},
       adsurl = {https://ui.adsabs.harvard.edu/abs/2025arXiv250315587S}
}

@ARTICLE{2023JOSS....8.5703P,
       author = {{Pasha}, Imad and {Miller}, Tim B.},
        title = "{pysersic: A Python package for determining galaxy structural properties via Bayesian inference, accelerated with jax}",
      journal = {The Journal of Open Source Software},
         year = 2023,
        month = sep,
       volume = {8},
       number = {89},
          eid = {5703},
        pages = {5703},
          doi = {10.21105/joss.05703},
archivePrefix = {arXiv},
       eprint = {2306.05454},
 primaryClass = {astro-ph.GA},
       adsurl = {https://ui.adsabs.harvard.edu/abs/2023JOSS....8.5703P}
}

@misc{Perrin2025,
  author    = {Perrin, M. and Long, J. and Osborne, S. and Geda, R. and Sappington, B. and Meléndez, M. and Lajoie, C.-P. and Leisenring, J. and Zimmerman, N. and Brooks, K. and Otor, O. J. and Kulp, T. and Chambers, L. and Jurling, A.},
  title     = {STPSF (2.1.0)},
  year      = {2025},
  howpublished = {Zenodo},
  doi       = {10.5281/zenodo.15747364},
  url       = {https://doi.org/10.5281/zenodo.15747364}
}

@ARTICLE{2018MNRAS.480.4379C,
       author = {{Carnall}, A.~C. and {McLure}, R.~J. and {Dunlop}, J.~S. and {Dav{\'e}}, R.},
        title = "{Inferring the star formation histories of massive quiescent galaxies with BAGPIPES: evidence for multiple quenching mechanisms}",
      journal = {\mnras},
         year = 2018,
        month = nov,
       volume = {480},
       number = {4},
        pages = {4379-4401},
          doi = {10.1093/mnras/sty2169},
archivePrefix = {arXiv},
       eprint = {1712.04452},
 primaryClass = {astro-ph.GA},
       adsurl = {https://ui.adsabs.harvard.edu/abs/2018MNRAS.480.4379C}
}

@ARTICLE{2019MNRAS.490..417C,
       author = {{Carnall}, A.~C. and {McLure}, R.~J. and {Dunlop}, J.~S. and {Cullen}, F. and {McLeod}, D.~J. and {Wild}, V. and {Johnson}, B.~D. and {Appleby}, S. and {Dav{\'e}}, R. and {Amorin}, R. and {Bolzonella}, M. and {Castellano}, M. and {Cimatti}, A. and {Cucciati}, O. and {Gargiulo}, A. and {Garilli}, B. and {Marchi}, F. and {Pentericci}, L. and {Pozzetti}, L. and {Schreiber}, C. and {Talia}, M. and {Zamorani}, G.},
        title = "{The VANDELS survey: the star-formation histories of massive quiescent galaxies at 1.0 < z < 1.3}",
      journal = {\mnras},
         year = 2019,
        month = nov,
       volume = {490},
       number = {1},
        pages = {417-439},
          doi = {10.1093/mnras/stz2544},
archivePrefix = {arXiv},
       eprint = {1903.11082},
 primaryClass = {astro-ph.GA},
       adsurl = {https://ui.adsabs.harvard.edu/abs/2019MNRAS.490..417C}
}

@Article{         harris2020array,
 title         = {Array programming with {NumPy}},
 author        = {Charles R. Harris and K. Jarrod Millman and St{\'{e}}fan J.
                 van der Walt and Ralf Gommers and Pauli Virtanen and David
                 Cournapeau and Eric Wieser and Julian Taylor and Sebastian
                 Berg and Nathaniel J. Smith and Robert Kern and Matti Picus
                 and Stephan Hoyer and Marten H. van Kerkwijk and Matthew
                 Brett and Allan Haldane and Jaime Fern{\'{a}}ndez del
                 R{\'{i}}o and Mark Wiebe and Pearu Peterson and Pierre
                 G{\'{e}}rard-Marchant and Kevin Sheppard and Tyler Reddy and
                 Warren Weckesser and Hameer Abbasi and Christoph Gohlke and
                 Travis E. Oliphant},
 year          = {2020},
 month         = sep,
 journal       = {Nature},
 volume        = {585},
 number        = {7825},
 pages         = {357--362},
 doi           = {10.1038/s41586-020-2649-2},
 publisher     = {Springer Science and Business Media {LLC}},
 url           = {https://doi.org/10.1038/s41586-020-2649-2}
}

@ARTICLE{2020SciPy-NMeth,
  author  = {Virtanen, Pauli and Gommers, Ralf and Oliphant, Travis E. and
            Haberland, Matt and Reddy, Tyler and Cournapeau, David and
            Burovski, Evgeni and Peterson, Pearu and Weckesser, Warren and
            Bright, Jonathan and {van der Walt}, St{\'e}fan J. and
            Brett, Matthew and Wilson, Joshua and Millman, K. Jarrod and
            Mayorov, Nikolay and Nelson, Andrew R. J. and Jones, Eric and
            Kern, Robert and Larson, Eric and Carey, C J and
            Polat, {\.I}lhan and Feng, Yu and Moore, Eric W. and
            {VanderPlas}, Jake and Laxalde, Denis and Perktold, Josef and
            Cimrman, Robert and Henriksen, Ian and Quintero, E. A. and
            Harris, Charles R. and Archibald, Anne M. and
            Ribeiro, Ant{\^o}nio H. and Pedregosa, Fabian and
            {van Mulbregt}, Paul and {SciPy 1.0 Contributors}},
  title   = {{{SciPy} 1.0: Fundamental Algorithms for Scientific
            Computing in Python}},
  journal = {Nature Methods},
  year    = {2020},
  volume  = {17},
  pages   = {261--272},
  adsurl  = {https://rdcu.be/b08Wh},
  doi     = {10.1038/s41592-019-0686-2},
}

@ARTICLE{2012MNRAS.426..296D,
       author = {{D{\'\i}az-Gim{\'e}nez}, Eugenia and {Mamon}, Gary A. and {Pacheco}, Marcela and {Mendes de Oliveira}, Claudia and {Alonso}, M. Victoria},
        title = "{Compact groups of galaxies selected by stellar mass: the 2MASS compact group catalogue}",
      journal = {\mnras},
         year = 2012,
        month = oct,
       volume = {426},
       number = {1},
        pages = {296-316},
          doi = {10.1111/j.1365-2966.2012.21705.x},
archivePrefix = {arXiv},
       eprint = {1207.2196},
 primaryClass = {astro-ph.CO},
       adsurl = {https://ui.adsabs.harvard.edu/abs/2012MNRAS.426..296D}
}

@ARTICLE{2025A&A...699A.329T,
       author = {{Tricottet}, Matthieu and {Mamon}, Gary A. and {D{\'\i}az-Gim{\'e}nez}, Eugenia},
        title = "{Are compact groups of galaxies special?}",
      journal = {\aap},
         year = 2025,
        month = jul,
       volume = {699},
          eid = {A329},
        pages = {A329},
          doi = {10.1051/0004-6361/202451713},
archivePrefix = {arXiv},
       eprint = {2407.20357},
 primaryClass = {astro-ph.GA},
       adsurl = {https://ui.adsabs.harvard.edu/abs/2025A&A...699A.329T}
}

@ARTICLE{1990AJ....100...32B,
       author = {{Beers}, Timothy C. and {Flynn}, Kevin and {Gebhardt}, Karl},
        title = "{Measures of Location and Scale for Velocities in Clusters of Galaxies---A Robust Approach}",
      journal = {\aj},
         year = 1990,
        month = jul,
       volume = {100},
        pages = {32},
          doi = {10.1086/115487},
       adsurl = {https://ui.adsabs.harvard.edu/abs/1990AJ....100...32B}
}

@ARTICLE{1985ApJ...298....8H,
       author = {{Heisler}, J. and {Tremaine}, S. and {Bahcall}, J.~N.},
        title = "{Estimating the masses of galaxy groups: alternatives to the virial theorem.}",
      journal = {\apj},
         year = 1985,
        month = nov,
       volume = {298},
        pages = {8-17},
          doi = {10.1086/163584},
       adsurl = {https://ui.adsabs.harvard.edu/abs/1985ApJ...298....8H}
}

@ARTICLE{1994ApJ...429..582C,
       author = {{Calzetti}, Daniela and {Kinney}, Anne L. and {Storchi-Bergmann}, Thaisa},
        title = "{Dust Extinction of the Stellar Continua in Starburst Galaxies: The Ultraviolet and Optical Extinction Law}",
      journal = {\apj},
         year = 1994,
        month = jul,
       volume = {429},
        pages = {582},
          doi = {10.1086/174346},
       adsurl = {https://ui.adsabs.harvard.edu/abs/1994ApJ...429..582C}
}

@ARTICLE{2020A&A...634A.135G,
       author = {{Girelli}, G. and {Pozzetti}, L. and {Bolzonella}, M. and {Giocoli}, C. and {Marulli}, F. and {Baldi}, M.},
        title = "{The stellar-to-halo mass relation over the past 12 Gyr. I. Standard {\ensuremath{\Lambda}}CDM model}",
      journal = {\aap},
         year = 2020,
        month = feb,
       volume = {634},
          eid = {A135},
        pages = {A135},
          doi = {10.1051/0004-6361/201936329},
archivePrefix = {arXiv},
       eprint = {2001.02230},
 primaryClass = {astro-ph.CO},
       adsurl = {https://ui.adsabs.harvard.edu/abs/2020A&A...634A.135G}
}

@ARTICLE{2025A&A...702A.163P,
       author = {{Paquereau}, L. and {Laigle}, C. and {McCracken}, H.~J. and {Shuntov}, M. and {Ilbert}, O. and {Akins}, H.~B. and {Allen}, N. and {Arango-Togo}, R. and {Berman}, E.~M. and {B{\'e}thermin}, M. and {Casey}, C.~M. and {McCleary}, J. and {Dubois}, Y. and {Drakos}, N.~E. and {Faisst}, A.~L. and {Franco}, M. and {Harish}, S. and {Jespersen}, C.~K. and {Kartaltepe}, J.~S. and {Koekemoer}, A.~M. and {Kokorev}, V. and {Lambrides}, E. and {Larson}, R. and {Liu}, D. and {Le Borgne}, D. and {Lewis}, J.~S.~W. and {McKinney}, J. and {Mercier}, W. and {Rhodes}, J.~D. and {Robertson}, B.~E. and {Toft}, S. and {Trebitsch}, M. and {Tresse}, L. and {Weaver}, J.~R.},
        title = "{Tracing the galaxy-halo connection with galaxy clustering in COSMOS-Web from z = 0.1 to z {\ensuremath{\sim}} 12}",
      journal = {\aap},
         year = 2025,
        month = oct,
       volume = {702},
          eid = {A163},
        pages = {A163},
          doi = {10.1051/0004-6361/202553828},
archivePrefix = {arXiv},
       eprint = {2501.11674},
 primaryClass = {astro-ph.GA},
       adsurl = {https://ui.adsabs.harvard.edu/abs/2025A&A...702A.163P}
}

@ARTICLE{2025ApJ...985...24C,
       author = {{Chakraborty}, Priyanka and {Sarkar}, Arnab and {Smith}, Randall and {Ferland}, Gary J. and {McDonald}, Michael and {Forman}, William and {Vogelsberger}, Mark and {Torrey}, Paul and {Garcia}, Alex M. and {Bautz}, Mark and {Foster}, Adam and {Miller}, Eric and {Grant}, Catherine},
        title = "{Unveiling the Cosmic Chemistry. II. ``Direct'' T$_{e}$-based Metallicity of Galaxies at 3 < z < 10 with JWST/NIRSpec}",
      journal = {\apj},
         year = 2025,
        month = may,
       volume = {985},
       number = {1},
          eid = {24},
        pages = {24},
          doi = {10.3847/1538-4357/adc7b5},
archivePrefix = {arXiv},
       eprint = {2412.15435},
 primaryClass = {astro-ph.GA},
       adsurl = {https://ui.adsabs.harvard.edu/abs/2025ApJ...985...24C}
}

@ARTICLE{2024ApJ...962...24S,
       author = {{Sanders}, Ryan L. and {Shapley}, Alice E. and {Topping}, Michael W. and {Reddy}, Naveen A. and {Brammer}, Gabriel B.},
        title = "{Direct T $_{e}$-based Metallicities of z = 2{\textendash}9 Galaxies with JWST/NIRSpec: Empirical Metallicity Calibrations Applicable from Reionization to Cosmic Noon}",
      journal = {\apj},
         year = 2024,
        month = feb,
       volume = {962},
       number = {1},
          eid = {24},
        pages = {24},
          doi = {10.3847/1538-4357/ad15fc},
archivePrefix = {arXiv},
       eprint = {2303.08149},
 primaryClass = {astro-ph.GA},
       adsurl = {https://ui.adsabs.harvard.edu/abs/2024ApJ...962...24S}
}

@ARTICLE{2023MNRAS.526.3504H,
       author = {{Hirschmann}, Michaela and {Charlot}, Stephane and {Somerville}, Rachel S.},
        title = "{High-redshift metallicity calibrations for JWST spectra: insights from line emission in cosmological simulations}",
      journal = {\mnras},
         year = 2023,
        month = dec,
       volume = {526},
       number = {3},
        pages = {3504-3518},
          doi = {10.1093/mnras/stad2745},
archivePrefix = {arXiv},
       eprint = {2305.03753},
 primaryClass = {astro-ph.GA},
       adsurl = {https://ui.adsabs.harvard.edu/abs/2023MNRAS.526.3504H}
}

@ARTICLE{2014MNRAS.444.3466S,
       author = {{Stanway}, Elizabeth R. and {Eldridge}, John J. and {Greis}, Stephanie M.~L. and {Davies}, Luke J.~M. and {Wilkins}, Stephen M. and {Bremer}, Malcolm N.},
        title = "{Interpreting high [O III]/H {\ensuremath{\beta}} ratios with maturing starbursts}",
      journal = {\mnras},
         year = 2014,
        month = nov,
       volume = {444},
       number = {4},
        pages = {3466-3472},
          doi = {10.1093/mnras/stu1682},
archivePrefix = {arXiv},
       eprint = {1408.4122},
 primaryClass = {astro-ph.GA},
       adsurl = {https://ui.adsabs.harvard.edu/abs/2014MNRAS.444.3466S}
}

@ARTICLE{2025ApJS..280...62L,
       author = {{Li}, Zihao and {Cai}, Zheng and {Wang}, Xin and {Li}, Zhaozhou and {Dekel}, Avishai and {Sarkar}, Kartick C. and {Ba{\~n}ados}, Eduardo and {Bian}, Fuyan and {Bhowmick}, Aklant K. and {Blecha}, Laura and {Bosman}, Sarah E.~I. and {Champagne}, Jaclyn B. and {Fan}, Xiaohui and {Golden-Marx}, Emmet and {Jun}, Hyunsung D. and {Li}, Mingyu and {Lin}, Xiaojing and {Liu}, Weizhe and {Sun}, Fengwu and {Trebitsch}, Maxime and {Walter}, Fabian and {Wang}, Feige and {Wu}, Yunjing and {Yang}, Jinyi and {Zhang}, Huanian and {Zhang}, Shiwu and {Zhuang}, Mingyang and {Zou}, Siwei},
        title = "{A 13 Billion Year View of Galaxy Growth: Metallicity Gradient Evolution from the Local Universe to z = 9 with JWST and Archival Surveys}",
      journal = {\apjs},
         year = 2025,
        month = oct,
       volume = {280},
       number = {2},
          eid = {62},
        pages = {62},
          doi = {10.3847/1538-4365/adfa70},
archivePrefix = {arXiv},
       eprint = {2506.12129},
 primaryClass = {astro-ph.GA},
       adsurl = {https://ui.adsabs.harvard.edu/abs/2025ApJS..280...62L}
}

@ARTICLE{2023ApJS..269...33N,
       author = {{Nakajima}, Kimihiko and {Ouchi}, Masami and {Isobe}, Yuki and {Harikane}, Yuichi and {Zhang}, Yechi and {Ono}, Yoshiaki and {Umeda}, Hiroya and {Oguri}, Masamune},
        title = "{JWST Census for the Mass-Metallicity Star Formation Relations at z = 4-10 with Self-consistent Flux Calibration and Proper Metallicity Calibrators}",
      journal = {\apjs},
         year = 2023,
        month = dec,
       volume = {269},
       number = {2},
          eid = {33},
        pages = {33},
          doi = {10.3847/1538-4365/acd556},
archivePrefix = {arXiv},
       eprint = {2301.12825},
 primaryClass = {astro-ph.GA},
       adsurl = {https://ui.adsabs.harvard.edu/abs/2023ApJS..269...33N}
}

@ARTICLE{2024ApJ...975..245C,
       author = {{Calabr{\`o}}, Antonello and {Castellano}, Marco and {Zavala}, Jorge A. and {Pentericci}, Laura and {Arrabal Haro}, Pablo and {Bakx}, Tom J.~L.~C. and {Burgarella}, Denis and {Casey}, Caitlin M. and {Dickinson}, Mark and {Finkelstein}, Steven L. and {Fontana}, Adriano and {Llerena}, Mario and {Mascia}, Sara and {Merlin}, Emiliano and {Mitsuhashi}, Ikki and {Napolitano}, Lorenzo and {Paris}, Diego and {P{\'e}rez-Gonz{\'a}lez}, Pablo G. and {Roberts-Borsani}, Guido and {Santini}, Paola and {Treu}, Tommaso and {Vanzella}, Eros},
        title = "{Evidence of Extreme Ionization Conditions and Low Metallicity in GHZ2/GLASS-Z12 from a Combined Analysis of NIRSpec and MIRI Observations}",
      journal = {\apj},
         year = 2024,
        month = nov,
       volume = {975},
       number = {2},
          eid = {245},
        pages = {245},
          doi = {10.3847/1538-4357/ad7602},
archivePrefix = {arXiv},
       eprint = {2403.12683},
 primaryClass = {astro-ph.GA},
       adsurl = {https://ui.adsabs.harvard.edu/abs/2024ApJ...975..245C}
}

@ARTICLE{2025A&A...700A..12M,
       author = {{Mazzolari}, Giovanni and {Scholtz}, Jan and {Maiolino}, Roberto and {Gilli}, Roberto and {Traina}, Alberto and {L{\'o}pez}, Ivan E. and {{\"U}bler}, Hannah and {Trefoloni}, Bartolomeo and {D'Eugenio}, Francesco and {Ji}, Xihan and {Mignoli}, Marco and {Vito}, Fabio and {Vignali}, Cristian and {Brusa}, Marcella},
        title = "{Narrow-line AGN selection in CEERS: Spectroscopic selection, physical properties, and X-ray and radio analysis}",
      journal = {\aap},
         year = 2025,
        month = aug,
       volume = {700},
          eid = {A12},
        pages = {A12},
          doi = {10.1051/0004-6361/202451860},
archivePrefix = {arXiv},
       eprint = {2408.15615},
 primaryClass = {astro-ph.GA},
       adsurl = {https://ui.adsabs.harvard.edu/abs/2025A&A...700A..12M}
}

@ARTICLE{2025A&A...697A.175S,
       author = {{Scholtz}, Jan and {Maiolino}, Roberto and {D'Eugenio}, Francesco and {Curtis-Lake}, Emma and {Carniani}, Stefano and {Charlot}, Stephane and {Curti}, Mirko and {Silcock}, Maddie S. and {Arribas}, Santiago and {Baker}, William and {Bhatawdekar}, Rachana and {Boyett}, Kristan and {Bunker}, Andrew J. and {Chevallard}, Jacopo and {Circosta}, Chiara and {Eisenstein}, Daniel J. and {Hainline}, Kevin and {Hausen}, Ryan and {Ji}, Xihan and {Ji}, Zhiyuan and {Johnson}, Benjamin D. and {Kumari}, Nimisha and {Looser}, Tobias J. and {Lyu}, Jianwei and {Maseda}, Michael V. and {Parlanti}, Eleonora and {Perna}, Michele and {Rieke}, Marcia and {Robertson}, Brant and {Del Pino}, Bruno Rodr{\'\i}guez and {Sun}, Fengwu and {Tacchella}, Sandro and {{\"U}bler}, Hannah and {Venturi}, Giacomo and {Williams}, Christina C. and {Willmer}, Christopher N.~A. and {Willott}, Chris and {Witstok}, Joris},
        title = "{JADES: A large population of obscured, narrow-line active galactic nuclei at high redshift}",
      journal = {\aap},
         year = 2025,
        month = may,
       volume = {697},
          eid = {A175},
        pages = {A175},
          doi = {10.1051/0004-6361/202348804},
archivePrefix = {arXiv},
       eprint = {2311.18731},
 primaryClass = {astro-ph.GA},
       adsurl = {https://ui.adsabs.harvard.edu/abs/2025A&A...697A.175S}
}

@ARTICLE{2025A&A...694A.218B,
       author = {{Brinch}, Malte and {Jin}, Shuowen and {Gobat}, Raphael and {Sillassen}, Nikolaj B. and {Algera}, Hiddo and {Gillman}, Steven and {Greve}, Thomas R. and {Gomez-Guijarro}, Carlos and {Gullberg}, Bitten and {Hodge}, Jacqueline and {Lee}, Minju and {Liu}, Daizhong and {Magdis}, Georgios and {Valentino}, Francesco},
        title = "{Revealing the hidden cosmic feast: A z = 4.3 galaxy group hosting two optically dark, efficiently star-forming galaxies}",
      journal = {\aap},
         year = 2025,
        month = feb,
       volume = {694},
          eid = {A218},
        pages = {A218},
          doi = {10.1051/0004-6361/202451448},
archivePrefix = {arXiv},
       eprint = {2501.05288},
 primaryClass = {astro-ph.GA},
       adsurl = {https://ui.adsabs.harvard.edu/abs/2025A&A...694A.218B}
}

@ARTICLE{2024A&A...683L...4J,
       author = {{Jin}, Shuowen and {Sillassen}, Nikolaj B. and {Magdis}, Georgios E. and {Brinch}, Malte and {Shuntov}, Marko and {Brammer}, Gabriel and {Gobat}, Raphael and {Valentino}, Francesco and {Carnall}, Adam C. and {Lee}, Minju and {Vijayan}, Aswin P. and {Gillman}, Steven and {Kokorev}, Vasily and {Le Bail}, Aur{\'e}lien and {Greve}, Thomas R. and {Gullberg}, Bitten and {Gould}, Katriona M.~L. and {Toft}, Sune},
        title = "{Cosmic Vine: A z = 3.44 large-scale structure hosting massive quiescent galaxies}",
      journal = {\aap},
         year = 2024,
        month = mar,
       volume = {683},
          eid = {L4},
        pages = {L4},
          doi = {10.1051/0004-6361/202348540},
archivePrefix = {arXiv},
       eprint = {2311.04867},
 primaryClass = {astro-ph.GA},
       adsurl = {https://ui.adsabs.harvard.edu/abs/2024A&A...683L...4J}
}

@software{sun2024nircam_grism,
  author       = {Sun, Fengwu},
  title        = {nircam\_grism},
  year         = {2024},
  month        = jul,
  version      = {3.0},
  publisher    = {Zenodo},
  doi          = {10.5281/zenodo.14052875},
  url          = {https://doi.org/10.5281/zenodo.14052875}
}
\bibliographystyle{aasjournalv7}



\end{document}